\documentclass[preprint,12pt]{elsarticle}

\usepackage{graphics}

\usepackage{amssymb}
\usepackage{color}

\journal{Chemical Physics}

\begin{document}

\begin{frontmatter}



\title{A microscopic mechanism for increasing\\
thermoelectric efficiency}

 \author[tokyo,jst]{Keiji Saito}
 \author[como,milano]{Giuliano Benenti}
 \author[como,milano,singapore]{Giulio Casati}
\address[tokyo]{Department of Physics, Graduate School of Science,
University of Tokyo, Tokyo 113-0033, Japan}
\address[jst]{Department of Physics, Graduate School of Science,
2 CREST, JST, 4-1-8 Honcho Kawaguchi, Saitama, 332-0012, Japan}
\address[como]{CNISM, CNR-INFM, and Center for Nonlinear and Complex Systems,\\
Universit\`a degli Studi dell'Insubria, Via Valleggio 11, 22100 Como, Italy}
\address[milano]{Istituto Nazionale di Fisica Nucleare, Sezione di Milano,\\
Via Celoria 16, 20133 Milano, Italy}
\address[singapore]{Centre for Quantum Technologies, 
National University of Singapore, Singapore 117543}

\begin{abstract}
We study the coupled particle and energy transport in a 
prototype model of interacting one-dimensional system: 
the disordered hard-point gas,
for which numerical data suggest that the thermoelectric 
figure of merit $ZT$ diverges with the system size. This result
is explained in terms of a microscopic mechanism, namely the local 
equilibrium is characterized by the emergence of a broad stationary 
``modified Maxwell-Boltzmann velocity distribution'', of width 
much larger than the mean velocity of the particle flow. 
\end{abstract}


\end{frontmatter}


\section{Introduction}
\label{sec:intro}

{Thermoelectricity is an old field: 
The Seebeck effect, that is, the conversion of temperature differences
into electricity, was discovered in 1821. 
However, a strong interest in termoelectric phenomena arose
only in the 1950's, when Ioffe discovered that doped 
semiconductors exhibited much 
larger thermoelectric effect than did other materials.
He also proposed that semiconductors could be used to 
build solid-state home refrigerators. Such refrigerators 
would be long-lived, silent, maintenace-free, and
environmentally benign. 
Ioffe's suggestion
initiated an intense research activity in semiconductors 
physics~\cite{mahanPT,majumdar,dresselhaus}. 
However, in spite of all efforts and consideration of all type  of
semiconductors, thermoelectric  refrigerators
have still poor efficiencies compared to compressor-based
refrigerators.  Today, thermoelectric devices  are mainly used in
situations in which reliability and quiet operation are
more important than the cost. Applications include 
equipments  in medical applications, space probes, etc.}  

In  the last  decade  there has  been  an increasing pressure  to  find
better thermoelectric  materials  with  higher efficiency.    The
reason   is   the  strong   environmental concern   about
chlorofluorocarbons  used in  most compressor-based refrigerators.
Also the possibility to  generate electric power from waste heat
using thermoelectric effect is becoming more and more 
interesting~\cite{mahanPT,majumdar,dresselhaus,snyder}.

The thermodynamic efficiency can be conveniently written in terms of the 
so-called figure of merit $ZT=(\sigma S^2/\kappa)T$, where $\sigma$
is the electric conductivity, $S$ the thermoelectric power 
(Seebeck coefficient), $\kappa$ the thermal conductivity, and $T$ the
temperature. Ideal Carnot efficiency is recovered in the limit 
$ZT\to\infty$. In spite of the worldwide research efforts for identifying
thermoelectric materials with high $ZT$ values, so far the best
thermoelectric materials are characterized
by values of $ZT\sim 1$, at room temperature.
Values $ZT>3$ are considered to be essential for thermoelectric 
devices to compete in efficiency with mechanical power generation and 
refrigeration. 

The challenge lies in engineering a material for which the values of 
$\sigma$, $S$, and $\kappa$ can be controlled in order to optimize 
thermoelectric efficiency. The problem is that the different transport 
coefficients are interdependent, thus making optimization extremely
difficult. On the other hand, thermodynamics does not impose any
upper bound on $ZT$, so that efficient thermoelectric devices could
in principle be engineered. 
The present understanding of the possible microscopic mechanisms leading
to an increase of $ZT$ is quite limited, with few exceptions. 
Notably, Refs.~\cite{mahan,linke} showed that the optimal density 
of states in a thermoelectric material is a delta function. Such 
sharp energy filtering allows to reach, in principle, the Carnot
efficiency. 

Here we consider the problem of increasing thermoelectric efficiency 
from a new perspective, that is, we pursue a dynamical 
system approach. 
{Understanding from first principles and from nonlinear dynamics 
simulations the microscopic mechanisms that can be implemented to control 
the heat flow~\cite{heatflow} might prove useful 
not only for thermoelectric phenomena but also for the 
design and engineering of thermal diodes and transistors.}
In this paper, our plan is to compute 
transport coefficients and thermoelectric efficiency 
from first principles, namely 
from the underlying microscopic 
dynamical processes which are known to be predominantly nonlinear in nature.
In a previous work~\cite{cas1}, the thermoelectric problem 
has been investigated by numerical solution of the microscopic equations of
motion. Inspired by the kinetic theory of ergodic gases and chaotic
billiards, a simple microscopic mechanism for 
increasing thermoelectric efficiency was proposed.  
More precisely, the cross transport of particles  and energy in open 
classical ergodic billiards was considered.  It has been shown that,
in  the linear
response regime, the thermoelectric efficiency can approach Carnot
efficiency for sufficiently complex charge carrier molecules. Indeed,
the figure of merit has been found to be a growing function of the
number $d_{\rm int}$ 
of internal degrees of freedom, $ZT = (d + 1 + d_{\rm int})/2$,
where $d$ is the geometric dimension.

In spite of the abstract nature of the model, the above paper opens
the possibility for a theoretical understanding of the basic
microscopic requirements that a classical dynamical system must
fulfill in order to lead to a high thermoelectric figure of merit. 
In particular,
the question arises whether inter-particle interaction might
increase the effective number of degrees of freedom, thus leading to
a higher figure of merit than in the noninteracting idealized $d$
dimensional gas, where $ZT=(d+1)/2$.
Alnog these lines a detailed numerical study of the cross heat and
particle transport has been performed for an open one-dimensional
disordered hard-point gas~\cite{ZThardpointgas}.
It has been found that $ZT$ diverges as
a power-law in thermodynamic limit, $ZT \propto N^b$, where $N$ is
the average number of particles in the system and $b\approx 0.79$.
Even though the above result could be, in principle, very
interesting, no indication was given concerning the microscopic mechanism
which is responsible for the increase of $ZT$. 
On the other hand a theoretical understanding is needed in
order to obtain useful hints for increasing thermodynamic efficiency
in more realistic models.

In this paper, we propose a mechanism which explains
the large thermoelectric quality factor $ZT$ numerically
observed in Ref.~\cite{ZThardpointgas}. This mechanism requires
local equilibrium, as naturally expected in systems with the 
mixing property, and the emergence, in the linear response regime, 
of an out-of-equilibrium ``modified Maxwell-Boltzmann velocity 
distribution'' of width much larger than the mean velocity of the 
particle flow. Such broad distribution limit is opposite to the 
limit of peaked distribution, corresponding to the  
delta-like energy filtering put forward
in Refs.~\cite{mahan,linke}. 
We provide numerical evidence supporting the effectiveness 
of the broad-distribution mechanism in the hard-point gas model.

Our paper is organized as follows. Secs.~\ref{sec:ZT},
\ref{sec:baths}, and \ref{sec:noninteracting} review 
introductory material on coupled particle and energy transport,
modeling stochastic baths, and thermoelectric efficiency 
of the one-dimensional ideal non-interacting gas.
In Sec.~\ref{sec:interacting}, we present numerical results 
for thermodynamic transport coefficients for the 
disordered hard-point gas model.
Finally, the obtained numerical results are explained in terms 
of a mechanism based on the emergence of a broad stationary
out-of-equilibrium velocity distribution.
Concluding remarks are drawn in Sec.~\ref{sec:conclusions}.

\section{The thermoelectric figure of merit $ZT$}
\label{sec:ZT}

Let us focus our attention on a conductor in which both
electric and heat current flow in one dimension
(say, parallel to the $x$-direction).
Assuming local equilibrium, a local entropy (per unit volume)
$s$ can be defined, and the rate of entropy production
reads~\cite{callen}
\begin{equation}
\dot{s}=J_u \partial_x \left(\frac{1}{T}\right) +
J_\rho \partial_x \left(-\frac{\mu}{T}\right),
\label{eq:dots}
\end{equation}
in which $J_u$ and $J_\rho$ are the energy and particle
current densities (fluxes) and $\partial_x (1/T)$,
$-\partial_x(\mu/T)$ the associated generalized forces
(affinities), where $T$ is the temperature and $\mu$
the electrochemical potential.

Assuming that the generalized forces are small, the
relationship between fluxes and forces is linear
and described by the phenomenological non equilibrium
thermodynamic kinetic equations~\cite{callen,mazur}
\begin{equation}
J_u = L_{uu} \partial_x \left(\frac{1}{T}\right)
     +L_{u\rho} \partial_x \left(-\frac{\mu}{T}\right),
\label{eq:Ju}
\end{equation}
\begin{equation}
J_\rho = L_{\rho u} \partial_x \left(\frac{1}{T}\right)
     +L_{\rho\rho} \partial_x \left(-\frac{\mu}{T}\right),
\label{eq:Jrho}
\end{equation}
with $L_{\alpha,\beta}$ ($\alpha,\beta\in\{u,\rho\}$) Onsager coefficients.
In the absence of magnetic fields, due to microscopic reversibility of
the dynamics, the Onsager reciprocity relation
$L_{u \rho}=L_{\rho u}$ holds.

In analogy with the relation $dQ=T dS$, the heat current density
$J_q$ can be defined by the relation
\begin{equation}
J_q= T J_s,
\end{equation}
with
\begin{equation}
J_s = \frac{1}{T} J_u - \frac{\mu}{T} J_\rho
\end{equation}
current density of entropy, and therefore
\begin{equation}
J_q=J_u - \mu J_\rho.
\end{equation}
The entropy production rate equation can then be written in terms of
the fluxes $J_q$ and $J_\rho$ and of the corresponding generalized forces
$\partial_x (1/T)$ and $-(1/T)\partial_x \mu$:
\begin{equation}
\dot{s}=J_q \partial_x \left(\frac{1}{T}\right) +
J_\rho \left(-\frac{\partial_x\mu}{T}\right),
\label{eq:dots2}
\end{equation}
while the linear relationship between fluxes and forces reads
as follows:
\begin{equation}
J_q = \tilde{L}_{qq} \partial_x \left(\frac{1}{T}\right)
     +\tilde{L}_{q\rho} \left(-\frac{\partial_x \mu}{T}\right),
\label{eq:Jq}
\end{equation}
\begin{equation}
J_\rho = \tilde{L}_{\rho q} \partial_x \left(\frac{1}{T}\right)
     +\tilde{L}_{\rho\rho} \left(-\frac{\partial_x \mu}{T}\right),
\label{eq:Jrho2}
\end{equation}
with $\tilde{L}_{\rho\rho}=L_{\rho\rho}$,
$\tilde{L}_{q\rho}=\tilde{L}_{\rho q}$ (Onsager relation),
$\tilde{L}_{q\rho}=L_{u\rho}-\mu L_{\rho\rho}$,
$\tilde{L}_{qq}=L_{uu}-2\mu L_{u\rho}+\mu^2 L_{\rho\rho}$.
Note that, if we call ${\bf L}$ and ${\bf \tilde{L}}$ the $2\times 2$
Onsager matrices with matrix elements $L_{\alpha \beta}$
($\alpha,\beta\in \{u,\rho\}$) and $\tilde{L}_{\gamma\delta}$
($\gamma,\delta\in \{q,\rho\}$), it turns out that
${\det}{\bf \tilde{L}}={\det}{\bf L}$.

The Onsager coefficients can be expressed in terms of more familiar
quantities, the electric conductivity $\sigma$, the thermal
conductivity $\kappa$, and the Seebeck coefficient (thermopower) $S$.
Let us first consider the case in which the thermal gradient vanishes,
$\partial_x T=0$, and the system is homogeneous, so that the chemical
potential $\mu_c$ is uniform. Since the electrochemical potential
$\mu$ is composed of a chemical part $\mu_c$ and an electric part
$\mu_e$, $\mu=\mu_c + \mu_e$, it turns out that for a homogeneous
isothermal system $\partial_x\mu=\partial_x \mu_e$.
The electric current $J_e=e J_\rho$, with $e$ charge of the
conducting particles, is then given by
$J_e=\sigma\mathcal{E}=-(\sigma/e)\partial_x \mu_e$, with
$\mathcal{E}$ external electric field applied to the system.
The quantities $\mu_c$ and $\mu_e$ cannot be determined separately by
the theory of irreversible thermodynamics~\cite{walstrom}:
only their combination
$\mu=\mu_c+\mu_e$ appears in the kinetic equations
(\ref{eq:Ju}) and (\ref{eq:Jrho}).
Based on this equivalence, we can write
$J_\rho=J_e/e=-(\sigma/e^2)\partial_x \mu$ even when $\mu_c\ne 0$,
provided $\partial_x T=0$, whence Eq.~(\ref{eq:Jrho2}) gives
\begin{equation}
\sigma=
\frac{e^2}{T} \tilde{L}_{\rho\rho}=
\frac{e^2}{T} L_{\rho\rho}.
\label{eq:sigma}
\end{equation}

The heat conductivity $\kappa$ is defined as the heat current
density per unit temperature gradient for zero electric current:
$J_q=-\kappa \partial_x T$, at $J_e=0$. Solving the two kinetic equations
(\ref{eq:Jq}) and (\ref{eq:Jrho2}) simultaneously, we obtain
\begin{equation}
\kappa=\frac{1}{T^2} \frac{\det \tilde{\bf L}}{\tilde{L}_{\rho\rho}}=
\frac{1}{T^2}\frac{\det {\bf L}}{{L}_{\rho\rho}}.
\label{eq:kappa}
\end{equation}

Finally, the Seebeck coefficient $S$ is defined as the change 
in electrochemical potential per unit charge, 
$-\partial_x\mu/e$, per unit change in temperature difference:
$S=-(1/e) \partial_x\mu/\partial_x T$, at $J_e=0$. We then obtain from
Eq.~(\ref{eq:Jrho2})
\begin{equation}
S=\frac{\tilde{L}_{q\rho}}{eT\tilde{L}_{\rho\rho}}=
\frac{1}{eT} \left( \frac{L_{u\rho}}{L_{\rho\rho}}-\mu\right).
\label{eq:Seebeck}
\end{equation}
It is of course possible to eliminate the three Onsager coefficients
$\tilde{L}_{qq}$, $\tilde{L}_{q\rho}$, and
$\tilde{L}_{\rho\rho}$ from the kinetic equations
(\ref{eq:Jq}) and (\ref{eq:Jrho2}), and rewrite such equations
is terms of the conductivities $\sigma$ and $\kappa$, and of the
thermopower $S$:
\begin{equation}
J_q = -(\kappa+T\sigma S^2) \partial_x T
      - \frac{T\sigma S}{e} \partial_x\mu,
\label{eq:Jq2}
\end{equation}
\begin{equation}
J_\rho = -\frac{\sigma}{e^2} \partial_x \mu
         - \frac{\sigma S}{e} \partial_x T.
\label{eq:Jrho3}
\end{equation}
By eliminating $\partial_x \mu$ from the above two equations
one can express $J_q$ in terms of $J_\rho$ and $\partial_x T$.
It is then easy to derive an interesting expression for the entropy
current density $J_s=J_q/T$~\cite{callen}:
\begin{equation}
J_s= e S J_\rho -\frac{\kappa}{T} \partial_x T,
\end{equation}
from which the Seebeck coefficient can be understood as the
entropy transported (per unit charge) by the electron flow.
The second contribution to the entropy flow, namely the term
$-(\kappa/T) \partial_x T$, is independent of the particle 
current.

The thermoelectric efficiency $\eta$, of converting the input heat into
output work, is determined by the non-dimensional figure of merit
\begin{equation}
ZT \equiv \frac{\sigma S^2}{\kappa} T. \label{exp_of_zt}
\end{equation}
To derive the relation between $\eta$ and $ZT$, we consider a
one-dimensional system whose left/right ends are connected with
left/right thermochemical reservoirs, with small temperature difference
$\Delta T \equiv T_R - T_L$
and electrochemical potential difference
$\Delta \mu \equiv \mu_R-\mu_L$.
The efficiency $\eta$ 
is given, under steady state conditions, by the ratio of the time
derivatives of the extracted work over the heat leaving the hot reservoir:
\begin{equation}
\eta=\frac{\dot{W}}{\dot{Q}}=\frac{\Delta \mu J_\rho}{J_q}.
\end{equation}
Using Eqs.~(\ref{eq:Jq2}) and (\ref{eq:Jrho3}) to eliminate
$\partial_x \mu$ and $J_q$, we obtain
\begin{equation}
\eta = \eta_C\,\frac{T}{\sigma \partial_x T}\,
\frac{J_e^2 + \sigma S \partial_x T J_e}{T S J_e - k \partial_x T},
\label{eq:etanoopt}
\end{equation}
where $\eta_C=1- T_R/T_L$ is the Carnot efficiency
(here we assume $T_L> T_R$).
The maximum efficiency for a given $\Delta T$ is derived after
optimizing (\ref{eq:etanoopt}) with respect to $J_e$:
\begin{equation}
\eta_{\rm max}=\eta_C \frac{\sqrt{ZT+1} -1}{\sqrt{ZT+1}+1}.
\end{equation}
The Carnot efficiency is therefore achieved in the limt $ZT\to\infty$.

Using Eqs.~(\ref{eq:sigma}), (\ref{eq:kappa}), and (\ref{eq:Seebeck}),
we can express $ZT$ in terms of the Onsager coefficients:
\begin{equation}
ZT=\frac{\tilde{L}_{q\rho}^2}{\det{\bf \tilde{L}}}=
\frac{(L_{u\rho}-\mu L_{\rho\rho})^2}{\det {\bf L}}.
\label{eq:ZTOnsager}
\end{equation}
The only thermodynamic restrictions to the Onsager coefficients come
from the positivity of the entropy production, $\dot{s}\geq 0$, which is
a quadratic form in the generalized forces $\partial_x (1/T)$ and
$-\partial_x (\mu/T)$ (see Eqs.~(\ref{eq:dots})-(\ref{eq:Jrho})) 
or $\partial_x (1/T)$ and $-(1/T)\partial_x \mu$ 
(see Eqs.~(\ref{eq:dots2})-(\ref{eq:Jrho2})).
Condition $\dot{s}\geq 0$ implies
$L_{uu},L_{\rho\rho}\geq 0$, $\det{\bf L}\ge 0$ in the first case,
$\tilde{L}_{qq},\tilde{L}_{\rho\rho}\geq 0$,
$\det{\bf \tilde{L}}\ge 0$ in the latter.
Thus, the only restriction to the thermoelectric figure of merit
is $ZT\geq 0$, so that in principle Carnot efficiency can be achieved.

It is clear from Eq.~(\ref{eq:ZTOnsager}) that $ZT$ diverges iff the
Onsager matrix ${\bf L}$ (or, equivalently, ${\bf \tilde{L}}$) is
ill-conditioned, that is, when the condition number
$\lambda_{1}({\bf L})/\lambda_{2}({\bf L})$ diverges,
where $\lambda_{1}({\bf L})$ and $\lambda_{2}({\bf L})$ are
the largest and the smallest eigenvalue of ${\bf L}$, 
respectively. The condition
number diverges iff the quantity
\begin{equation}
{\rm cond} ({\bf L}) \equiv
\frac{[{\rm Tr} ({\bf L})]^2}{\det({\bf L})}
\label{eq:condnumber}
\end{equation}
diverges. In this case the system (\ref{eq:Ju})-(\ref{eq:Jrho})
(or, equivalently, the system (\ref{eq:Jq})-(\ref{eq:Jrho2})) becomes
singular, and therefore $J_u \propto J_\rho$. In short, the Carnot
efficiency is obtained iff the energy and particle currents
are proportional.

\section{Modeling thermochemical baths}
\label{sec:baths}

We consider a one-dimensional system whose ends are in contact with
left/right baths (reservoirs), which are able to exchange energy and particles
with the system, at fixed temperature $T_\alpha$
and electrochemical potential $\mu_\alpha$, where $\alpha=L,R$
denotes the left/right bath.

The thermochemical reservoirs are modeled as infinite
one-dimensional ideal gases. Therefore, particle velocities
in the reservoirs are described by the Maxwell-Boltzmann distribution,
\begin{equation}
f_\alpha(v)=\sqrt{\frac{m}{2\pi k_B T_\alpha}}\exp\left(
-\frac{m v^2}{2 k_B T_\alpha}\right),
\label{eq:MB}
\end{equation}
where $k_B$ is the Boltzmann constant and $m$ the mass of
the particles. We use a stochastic model of the thermochemical
baths~\cite{carlos}: Whenever a particle of the system crosses
the boundary which separates the system
 from the left or right reservoir,
it is removed. On the other hand, particles are injected into the
system from the boundaries, with rates $\gamma_\alpha$.
The injection rate $\gamma_\alpha$ is computed by counting
how many particle from reservoir $\alpha$ cross the
reservoir-system boundary per unit time. That is to say,
\begin{equation}
\gamma_\alpha=\rho_\alpha \int_0^\infty dv v f_\alpha(v)
= \rho_\alpha \sqrt{\frac{k_B T_\alpha}{2\pi m}},
\label{eq:num1}
\end{equation}
with $\rho_\alpha$ density of the ideal gas in reservoir
$\alpha$. Therefore, particles are injected into the system
with velocity distribution
\begin{equation}
P_\alpha(v)=\frac{m}{k_B T_\alpha} \, v \exp\left(
-\frac{m v^2}{2 k_B T_\alpha}\right)\theta_\alpha(v),
\label{eq:num2}
\end{equation}
where $\theta_\alpha(v)$ are step functions:
$\theta_L(v)=1$ if $v\ge 0$, 0 otherwise;
$\theta_R(v)=1$ if $v\le 0$, 0 otherwise.
{We assume that injections from a macroscopic reservoir are
independent events and that the time interval between subsequent 
injections satisfies the Poissonian distribution,}
\begin{equation}
{\cal P}_\alpha(t)=\gamma_\alpha \exp(-\gamma_\alpha t),
\label{eq:num3}
\end{equation}
so that the average time between injections is $1/\gamma_\alpha$.

In order to relate the density $\rho_\alpha$ to the 
electrochemical potential
$\mu_\alpha$, it is convenient to write the grand partition function
\begin{equation}
\Xi_\alpha=\sum_{N=0}^\infty \frac{1}{N !}
\left\{ \frac{\Lambda}{h} e^{\beta_\alpha \mu_\alpha}
\int dv \, m \exp \left[-\beta_\alpha \left(
\frac{1}{2} m v^2\right)\right]\right\}^N,
\label{eq:grandcanonical}
\end{equation}
with ${\Lambda}$ and $N$ size and number of particles
of the reservoir, respectively~\footnote{It is of course understood 
that ${\Lambda}$
is macroscopically large and that the thermodynamic limit is
eventually taken for the reservoir}, $\beta_\alpha\equiv 1/(k_B T_\alpha)$
and $h$ the Planck's constant.
We then compute the average
number of particles as
\begin{equation}
\langle N \rangle_\alpha = \frac{1}{\beta_\alpha}\frac{\partial}{\partial
\mu_\alpha} \ln \Xi_\alpha,
\end{equation}
so that
\begin{equation}
\rho_\alpha=\frac{\langle N \rangle_\alpha}{{\Lambda}}=
\frac{e^{\beta_\alpha \mu_\alpha}\sqrt{2\pi m k_B T_\alpha}}{h}.
\label{eq:num5}
\end{equation}
Therefore, we can express the 
electrochemical potentials of the bath in terms
of the injection rates:
\begin{equation}
\mu_\alpha = k_B T_\alpha \ln (\lambda_{T_\alpha} \rho_\alpha),
\label{eq:num4}
\end{equation}
with
\begin{equation}
\lambda_{T_\alpha}=\frac{h}{\sqrt{2\pi m k_B T_\alpha}}
\end{equation}
de Broglie thermal wave length.
Note that this relation, even though derived from the grand
partition function of a classical ideal gas, can only be justified
if particles are considered as indistinguishable. The
$1/N!$ term in the grand partition function (\ref{eq:grandcanonical})
is rooted in the above indistinguishability,
of purely quantum mechanical origin~\cite{huang}.
The stochastic thermochemical baths used in our numerical
simulations are based on
Eqs.~(\ref{eq:num1}), (\ref{eq:num2}), (\ref{eq:num3}),
and (\ref{eq:num4}). The 
electrochemical potential $\mu_\alpha$
and the temperature $T_\alpha$ can be controlled by varying
the injection rate $\gamma_\alpha$ and the temperature $T_\alpha$.

\section{One-dimensional non-interacting classical gas}
\label{sec:noninteracting}

Let us first consider the simplest case of a one-dimensional gas 
of non-interacting particles.
Assuming that also the reservoirs are one-dimensional and that the
left/right contacts between system and reservoirs are identical and
described as in Sec.~\ref{sec:baths}, the particle current $J_\rho$ reads
\begin{equation}
J_\rho=
\gamma_L\int_0^\infty d\epsilon u_L(\epsilon) {\cal T} (\epsilon)
-\gamma_R\int_0^\infty d\epsilon u_R(\epsilon) {\cal T} (\epsilon),
\label{eq:nonint1}
\end{equation}
where $u_\alpha(\epsilon)$ is the energy distribution of the
particles injected from reservoir $\alpha$
and ${\cal T}(\epsilon)$ is the transmission
probability for a particle with energy $\epsilon$ to transit from
one end to the other end of the system, $0\le {\cal T}(\epsilon)\le 1$.
Using Eq.~(\ref{eq:num2}), we obtain
\begin{equation}
u_\alpha(\epsilon)=\beta_\alpha e^{-\beta_\alpha \epsilon}.
\label{eq:nonint2}
\end{equation}
Furthermore, from Eqs.~(\ref{eq:num1}) and (\ref{eq:num5}) we have
\begin{equation}
\gamma_\alpha =
\frac{1}{h\beta_\alpha} e^{\beta_\alpha \mu_\alpha}.
\label{eq:nonint3}
\end{equation}
After substitution of (\ref{eq:nonint2}) and
(\ref{eq:nonint3}) into (\ref{eq:nonint1}), we arrive to the
following expression for the particle current:
\begin{equation}
J_\rho=\frac{1}{h}\int_0^\infty d\epsilon
\left( e^{-\beta_L (\epsilon-\mu_L)}-
e^{-\beta_R (\epsilon-\mu_R)} \right) {\cal T}(\epsilon).
\label{eq:jrhodelta}
\end{equation}
Similarly, we obtain the heat currents 
$J_{q,\alpha}=J_u-\mu_\alpha J_\rho$ at the left and right reservoirs:
\begin{equation}
J_{q,\alpha}=\frac{1}{h}\int_0^\infty d\epsilon (\epsilon - \mu_\alpha)
\left( e^{-\beta_L (\epsilon-\mu_L)}-
e^{-\beta_R (\epsilon-\mu_R)} \right) {\cal T}(\epsilon).
\end{equation}
The thermoelectric efficiency is then given by
(we assume $T_L>T_R$, $\mu_R>\mu_L$ and consider only
${\cal T}(\epsilon)$ functions such that $J_\rho\ge 0$ and
$J_{q,L}\ge 0$)
\begin{equation}
\eta=\frac{J_{q,L}-J_{q,R}}{J_{q,L}}=
\frac{(\mu_R-\mu_L)\int_0^\infty d\epsilon
\left( e^{-\beta_L (\epsilon-\mu_L)}-
e^{-\beta_R (\epsilon-\mu_R)} \right)
{\cal T}(\epsilon)}{\int_0^\infty d\epsilon (\epsilon - \mu_L)
\left( e^{-\beta_L (\epsilon-\mu_L)}-
e^{-\beta_R (\epsilon-\mu_R)} \right) {\cal T}(\epsilon)}.
\end{equation}
When the transmission is possible only within a tiny energy 
window around $\epsilon=\epsilon_\star$, the efficiency reads
\begin{equation}
\eta= \frac{\mu_R-\mu_L}{\epsilon_\star-\mu_L}.
\label{eq:etalinke}
\end{equation}
In the limit $J_\rho\to 0$, corresponding to reversible 
transport~\cite{linke}, we get $\epsilon_\star$ 
from Eq.~(\ref{eq:jrhodelta}): 
\begin{equation}
\epsilon_\star=\frac{\beta_L\mu_L-\beta_R\mu_R}{\beta_L-\beta_R}.
\label{eq:etalinke2}
\end{equation}
Substituting such $\epsilon_\star$ in Eq.~(\ref{eq:etalinke}),
we obtain the Carnot efficiency $\eta=\eta_C=1-T_R/T_L$.
Such delta-like energy-filtering
mechanism for increasing thermoelectric efficiency
has been pointed out in Refs.~\cite{mahan,linke}.

In the linear response regime, using a delta-like energy filtering,
${\cal T}(\epsilon)=1$ in a tiny interval of width $\delta \epsilon$ 
{around 
some energy $\bar{\epsilon}$, $0$ otherwise, we obtain
\begin{equation}
L_{uu} = {L \bar{\epsilon}^2 (\delta \epsilon)\over hk_B} e^{-\beta 
(\bar{\epsilon} -\mu )}  , \;
L_{u\rho} =  L_{\rho u}=
{L \bar{\epsilon} 
(\delta \epsilon)\over hk_B} e^{-\beta (\bar{\epsilon} -\mu )}  ,  \;
L_{\rho \rho} = {L(\delta\epsilon) \over hk_B} 
e^{-\beta (\bar{\epsilon} -\mu )}  ,
\label{eq:deltalinear}
\end{equation}
where $L$ is the length of system.
From these relations we immediately derive that the Onsager matrix 
is ill-conditioned and therefore $ZT=\infty$ and
$\eta=\eta_C$.
We point out that the parameters $\bar{\epsilon}$ and
$\delta \epsilon$ characterizing the transmission window,
appear in the Onsager matrix elements (\ref{eq:deltalinear}) 
and therefore are assumed to be 
independent of the applied temperature and 
electrochemical potential 
gradients. On the other hand, the energy 
$\epsilon_\star$ in Eqs.~(\ref{eq:etalinke}),(\ref{eq:etalinke2})
depends on the applied gradients. 
There is of course no contradiction since
(\ref{eq:etalinke}),(\ref{eq:etalinke2}) have general validity 
beyond the linear response regime.} 

\section{One-dimensional interacting classical gas}
\label{sec:interacting}

\begin{figure}
\vspace{1cm}
\centering
\includegraphics[width=3.1in]{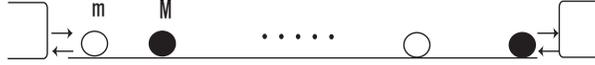}
\caption{Schematic picture of the model considered in
our numerical simulations.}
\label{schematic}
\end{figure}

Let us now turn to the interacting case. 
We consider a one-dimensional, di-atomic
disordered chain, of hard-point elastic particles with coordinates
$x_i\in [0,L]$, $L$ being the system size, velocities $v_i$ and
masses $m_i\in \{m,M\}$ randomly distributed. The particles interact
among themselves through elastic collisions only. 
A schematic picture of the model is drawn in Fig.~\ref{schematic}.
{Since we are considering a purely 
mechanical model, strictly speaking we are going to investigate
thermodiffusion rather than thermoelectricity. On the other hand, 
we assume that the particles are charged and that the 
Coulomb repulsion is screened and modeled
by a short-range hard-core interaction (elastic collisions). 
Therefore, our model is relevant also for thermoelectricity.}
Numerical results obtained in Ref.~\cite{ZThardpointgas} suggest
that, for mass ratio $M/m\ne 1$, the figure of merit $ZT$ diverges
in the thermodynamic limit.~\footnote{{The two masses must be
different in order to have ergodic and mixing dynamics, 
so that thermalization within the system occurs. 
For equal masses the dynamics is integrable 
and $ZT=1$~\cite{ZThardpointgas}.}}

Let $\ell$ be a reference unit length which we take $1$ in simulations.
In our numerical simulations we set
$\mu=(\mu_L+\mu_R)/2=0.2\,[h^2/m\ell^2]$ and 
$T=(T_L+T_R)/2=3.0\,[h^2/m\ell^2 k_B]$, and consider 
$\mu_L$, $\mu_R$, $T_L$, $T_R$
slightly
different from the mean values $\mu$, $T$ to drive finite 
currents $J_\rho$ and $J_u$.
{We assume that the mass of each particle 
injected by the left or right bath is 
chosen randomly and with equal a priori probabilities 
between the two possible values $m$ and $M$.}
The average currents $J_u$ and $J_{\rho}$ are computed  
at the contacts between system and baths:
If, in a period of time $t$ the left bath injects $N_i$ particles
with masses $m_j^{(i)}$
and velocities $v_j^{(i)}$, 
$j=1,...,N_i$, and absorbs $N_a$ particles with masses $m_j^{(a)}$
and velocities $v_j^{(a)}$, $j=1,...,N_a$, then in the large $t$
limit the currents $J_\rho$ and $J_u$ are given by 
\begin{equation}
J_\rho=\frac{1}{t} (N_i-N_a),
\end{equation}
\begin{equation}
J_u=\frac{1}{t}
\left(
\sum_{j=1}^{N_i} \frac{1}{2} m_j^{(i)} [v_j^{(i)}]^2
-\sum_{j=1}^{N_a} \frac{1}{2} m_j^{(a)} [v_j^{(a)}]^2
\right).
\end{equation}
Note that in the steady state, due to particle and energy conservation, 
these currents are equal to the corresponding currents computed for the 
right bath. 
Then the Onsager matrix elements from which $\sigma$, $S$, $\kappa$,
and $ZT$ can be readily derived, are obtained from 
Eqs.~(\ref{eq:Jq}) and (\ref{eq:Jrho2}). 
We set the mass ration $M/m=\pi$
and calculate currents up to $L=80 [\ell]$,
corresponding to an average number of particles 
inside the system $\langle N \rangle \approx 515$. 

\begin{figure}
\vspace{1cm}
\centering
\includegraphics[width=3.1in]{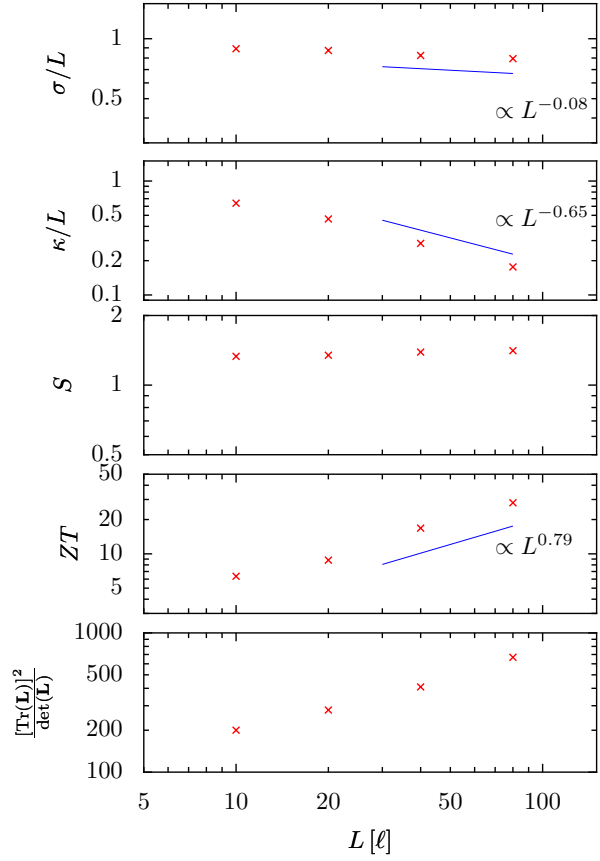}
\caption{Thermoelectric transport 
properties.
The quantities
$\sigma /L, \kappa /L$, $ZT$, and 
$\frac{[{\rm Tr} ({\bf L})]^2}{\det({\bf L})}$
show a power 
law dependence on the system size $L$.}
\label{zt_etc}
\end{figure}

In Fig.~\ref{zt_etc}, we present our numerical results 
for the transport coefficients. 
A power law dependence for $\sigma /L$, $\kappa /L$, and $ZT$ is 
observed~\footnote{Numerical data are consistent with 
those reported in Ref.~\cite{ZThardpointgas} for the same mass ratio.}.
In particular, the figure of merit $ZT$ increases with increasing
the systems size, 
$ZT \propto L^{0.79}$.
Correspondingly, the condition number
${[{\rm Tr} ({\bf L})]^2}/{\det({\bf L})}$ 
(see Eq.~(\ref{eq:condnumber})) diverges, as expected from 
the general theoretical considerations of Sec.~\ref{sec:ZT}.

\begin{figure}
\vspace{1cm}
\centering
\includegraphics[width=3.2in]{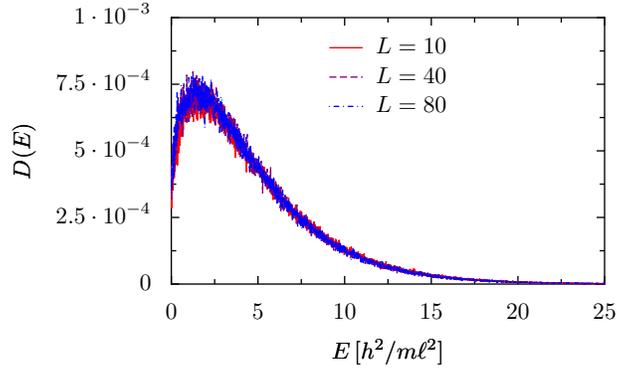}
\caption{$D(E)$ calculated for $L=10,40$ and $80$.}
\label{de}
\end{figure}

These numerical results naturally raise a question: Is the
mechanism leading to high $ZT$ quality factor for interacting
gases related to the delta-like mechanism~\cite{mahan,linke}
shortly discussed
in Sec.~\ref{sec:noninteracting} for the non-interacting ideal gas?
To address this question, we measure the particle current at
the position $x\in [0,L]$ as
\begin{equation}
J_\rho= \int_0^\infty dE D(E),
\end{equation}
\begin{equation}
D(E)\equiv D_L(E)-D_R(E),
\end{equation}
where the ``transmission function''
$D_L(E)$ is the density of particles with energy $E$
crossing $x$ and coming from the left side, while $D_R(E)$
is the density of particles with energy $E$ from the right side.
We inquire how $D(E)$ changes as a function of $L$, in particular
if $D(E)$ becomes more and more delta-like (peaked in energy)
when increasing $L$. The transmission function $D(E)$ is shown
in Fig.~\ref{de}, at $x=L/2$ and for different system 
sizes~\footnote{Note that,
while $J_\rho$ is position-independent due to conservation
of particles, $D(E)$ depends on $x$. However, we have checked that
similar behaviors of $D(E)$ are obtained for different values
of $x$.}.
There is no sign of narrowing of $D(E)$ when increasing the 
system size.
We can therefore conclude that the
mechanism leading to the large $ZT$ values observed
in Fig.~\ref{zt_etc} must be different
from the energy filtering discussed in Refs.~\cite{mahan,linke}.

To understand the mechanism,
we first write the particle and energy currents as
\begin{equation}
J_\rho= \overline{v(x,t)\, \rho(x,t)}
 ,
\quad
J_u =  \overline{\frac{1}{2} m v(x,t)^3 \rho (x,t)} ,
\label{eq:jvrho}
\end{equation}
where $x\in [0,L]$, the overbar denotes time-averaging, and
$v(x,t)$, $\rho (x,t)$ are respectively
the particles velocity and density at the position
$x$ and time $t$. 
If the relaxation time scales for density and velocity are well separated,
then expressions (\ref{eq:jvrho}) can be approximated as:
\begin{equation}
 J_\rho \sim \overline{v(x,t)} \times \overline{\rho(x,t)} , ~~~~~
 J_u \sim \overline{\frac{1}{2} m v(x,t)^3 } \times \overline{\rho (x,t)} \label{appro12}.
\label{eq:jrhouapprox}
\end{equation}
In our model, this is satisfied. For instance, in the case of 
$(\mu_L ,\mu_R)=(0.24,0.16)\, [h^2/m\ell^2], T=3.0\, [h^2/m\ell^2k_B]$, and
$L=40\, [\ell]$, we get at $x=L/2$ the time-averaged
velocity $\overline {v (L/2, t)}\approx 0.010\, [h/m\ell]$, and the time-averaged density
$\overline{\rho ( L/2,t)}\approx 6.43\, [1/\ell]$,
while $\overline{v(L/2,t)\, \rho(L/2,t)}
\approx 0.0641 \, [h/m\ell^2]$.

\begin{figure}
\vspace{1cm}
\centering
\includegraphics[width=3.2in]{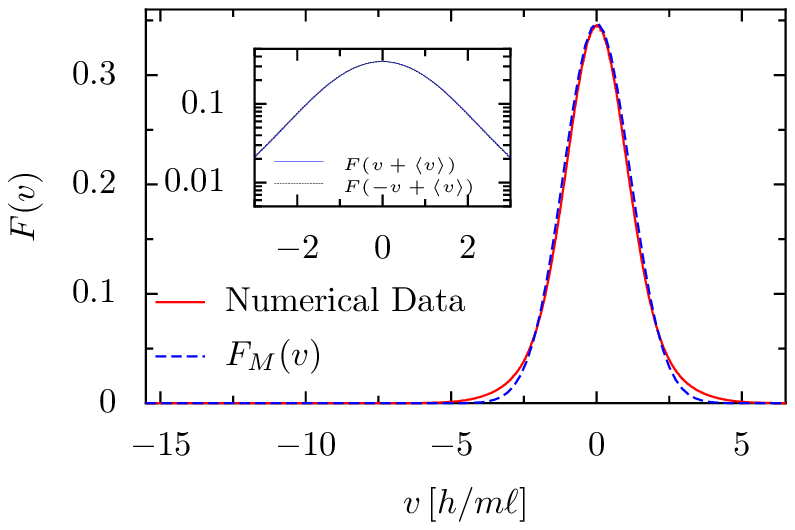}
\caption{
$F(v)$ calculated for the parameters $(\mu_L
 ,\mu_R)=(0.24,0.16) \, [h^2/m\ell^2 ], T=3.0\, [h^2/m\ell^2k_B]$ 
with the system size $L=40$. The solid line corresponds  
to the the numerical data, at $x=L/2$. 
The mean velocity is $\overline {v (L/2, t)}\sim 0.010\, [h/m\ell]$.
The dashed line is the modified Maxwell-Boltzmann distribution $F_M (v)$ 
which fits $F(v)$ with the parameters 
$\langle v \rangle = \overline {v (L/2, t)}= 0.010 [h/m\ell]$ and 
$\nu=1.15[h/m\ell]$. For these parameters, Eq.(\ref{v3}) yields
$\langle v^3 \rangle \sim 0.041$, which is comparable to the exact
 numerical data $\langle v^3 \rangle \sim 0.047$.  
The inset shows the behavior of $F(v+\langle v \rangle )$ and
 $F(-v+\langle v \rangle)$ 
in the semi-log scale. These two curves
completely overlap each other, as expected from 
Eq.~(\ref{eq:MBmod}).
Note that the tails deviate from the Gaussian form.
}
\label{ve}
\end{figure}

From the discussion of Sec.~\ref{sec:ZT}, it is clear that $ZT$
diverges when $J_u \propto J_\rho$. 
According to Eq.~(\ref{eq:jrhouapprox}), this is the case when
$\overline{v^3} \propto \overline{v}$.
Since we are interested in the steady-state transport properties
and we are considering systems with the mixing property, 
it is natural to assume that the time-averages
$\overline{v^n}$ equal the ensemble averages
$\langle v^n \rangle\equiv \int_{-\infty}^{+\infty} dv v^n F(v)$, with $F(v)$
velocity distribution function for the steady state.
At equilibrium ($T_L=T_R$, $\mu_L=\mu_R$), the system thermalizes
and $F(v)$ is the Maxwell-Boltzmann distribution
(\ref{eq:MB}) at any $x$. In the linear response regime, we
assume that $F(v)$ is given by a ``modified Maxwell-Boltzmann
distribution'',
\begin{equation}
F_M (v)=\sqrt{\frac{m^\star}{2\pi k_B T}}\exp\left(
-\frac{m^\star (v-\langle v \rangle)^2}{2 k_B T}\right),
\label{eq:MBmod}
\end{equation}
where the mean velocity $\langle v \rangle$ and 
the effective mass  $m^\star$ are fitting
parameters, and $T\approx T_L\approx T_R$.
That is to say, we assume that the out-of-equilibrium stationary
distribution (\ref{eq:MBmod}) differs from the equilibrium
Maxwell-Boltzmann distribution only in the position $\langle v \rangle$
of the peak, while the Gaussian shape is unchanged.
As shown in Fig.~\ref{ve}, such assumption is in good agreement with the
numerically computed $F(v)$ close to the peak of the distribution, while
the tails show deviations from (\ref{eq:MBmod}). Nevertheless,
such deviations do not affect too much the values of
$\langle v \rangle$ and $\langle v^3 \rangle$
and Eq.~(\ref{eq:MBmod}) is very convenient for analytical considerations
and to unveil the mechanism at the origin of the large thermoelectric
efficiencies observed in the hard-point gas model.

From Eq.~(\ref{eq:MBmod}) we obtain
\begin{equation}
\langle v^3 \rangle = \langle v \rangle^3 + 3 \nu^2 \langle v
 \rangle,
\label{v3}
\end{equation}
where
\begin{equation}
\nu \equiv \sqrt{\frac{k_B T}{m^\star}}
\end{equation}
is the width of distribution (\ref{eq:MBmod}). 
{We obtain
$\langle v^3 \rangle \propto \langle v \rangle$
when $\nu \gg \langle v \rangle$, that is, 
in the broad-distribution limit.
It is clear from Fig.~\ref{ve} that, for
the one-dimensional interacting hard-point gas,
indeed $\nu \gg \langle v \rangle$.~\footnote{
Note that the delta-like limit of Eq.~(\ref{eq:MBmod}),
$\nu\ll \langle v \rangle$, is incompatible with 
the linear response regime 
plus approximation (\ref{eq:jrhouapprox}) since, 
if $J_\rho\propto \langle v \rangle$ is a linear function
of the applied temperature and electrochemical potential gradients,
the same cannot hold for $J_u\propto \langle v^3 \rangle
\approx \langle v \rangle^3$.}} 

\section{Conclusions}
\label{sec:conclusions}

We have studied numerically the coupled particle and energy transport 
in a prototype model of interacting one-dimensional gas: the 
disordered, hard-point gas. 
There is numerical evidence that the $ZT$ quality factor diverges with increasing
the system size. We explain this result in terms of the emergence of a broad 
velocity distribution of the particles transmitted across the sample. 
This mechanism first of all requires local equilibrium, 
which is expected to take place
in systems with the mixing property. 
We also make a couple of assumptions which are quite 
natural in many-body systems:
the separation of the relaxation time scales of density and velocity in
Eq.~(\ref{appro12}), and the modified Maxwell-Boltzmann form of the 
velocity distribution (\ref{eq:MBmod}). 
On the other hand, since $ZT=(\sigma S^2/\kappa)T$ and 
Fig.~\ref{zt_etc} shows that the Seebeck coefficient is practically 
constant, the 
anomalous behavior of $\sigma$ and $\kappa$~\cite{lepri} 
is crucial to obtain a diverging $ZT$.
The relationship between the broad velocity-distribution mechanism 
and the anomalous behavior of the transport coefficients 
must be clarified. In particular, further 
investigations are required to understand whether this mechanism 
could be applied to systems with the mixing property but 
without anomalous transport. 
It might indeed be possible to find systems in which $\sigma$, $\kappa$
and $ZT$ eventually converge to finite but large values, when increasing
the system size.   
Therefore, our mechanism could be also relevant
in more realistic interacting systems with the mixing property. 

\section*{Acknowledgements}

G.B. and G.C. acknowledge support by the MIUR-PRIN 2008
{\it Efficiency of thermoelectric machines: A microscopic approach}.






\begin{thebibliography}{00}


\bibitem{mahanPT} 
G. Mahan, B. Sales, J. Sharp, Phys. Today  50 (March 1997), 42.

\bibitem{majumdar} 
A. Majumdar, Science 303 (2004) 777.

\bibitem{dresselhaus} 
M.S. Dresselhaus, G. Chen, M.Y. Tang, R.G. Yang, H. Lee, 
D.Z. Wang, Z.F. Ren, J.-P. Fleurial, P. Gogna,
Adv. Mater. 19 (2007) 1043.

\bibitem{snyder}
G.J. Snyder, E.R. Toberer, 
Nature Materials 7 (2008) 105.

\bibitem{mahan}
G.D. Mahan, J.O. Sofo,
Proc. Natl. Acad. Sci. USA 93 (1996) 7436.

\bibitem{linke}
T.E. Humphrey, R. Newbury, R.P. Taylor, H. Linke,
Phys. Rev. Lett. 89 (2002) 116801;
T.E. Humphrey, H. Linke,
Phys. Rev. Lett. 94 (2005) 096601.

{
\bibitem{heatflow}
M. Terraneo, M. Peyrard, G. Casati, 
Phys. Rev. Lett. 88 (2002) 094302;
B. Li, L. Wang, G. Casati, 
Phys. Rev. Lett. 93 (2004) 184301; 
D. Segal, A. Nitzan, 
Phys. Rev. Lett. 94 (2005) 034301;
B. Hu, L. Yang, and Y. Zhang, 
Phys. Rev. Lett. 97 (2006) 124302;
N. Yang, N. Li, L. Wang, B. Li, 
Phys. Rev. B 76, 020301(R) (2007);
B. Li, L. Wang, G. Casati, 
Appl. Phys. Lett. 88 (2006) 143501;
L. Wang and B. Li, 
Phys. World 21 (2008) 27;
N. Li, F. Zhan, P. H\"anggi, B. Li,
Phys. Rev. E 80 (2009) 011125,
and references therein.
}

\bibitem{cas1} 
G. Casati, C. Mej\'{\i}a-Monasterio, T. Prosen,
Phys. Rev. Lett. 101 (2008) 016601.

\bibitem{ZThardpointgas}
G. Casati, L. Wang, T. Prosen,
J. Stat. Mech. (2009) L03004.

\bibitem{callen}
H.B. Callen, Thermodynamics and an Introduction to Thermostatics
(second edition), John Wiley \& Sons, New York, 1985.

\bibitem{mazur}
S.R. de Groot, P. Mazur, Non-Equilibrium Thermodynamics,
Dover, New York, 1984.

\bibitem{walstrom}
P.L. Walstrom,
Am. J. Phys. 56 (1988) 890.

\bibitem{carlos}
C. Mej\'{\i}a-Monasterio, H. Larralde, F. Leyvraz,
Phys. Rev. Lett. 86 (2001) 5417;
H. Larralde, F. Leyvraz, C. Mej\'{\i}a-Monasterio,
J. Stat. Phys.  113 (2003) 197.

\bibitem{huang}
K. Huang, Statistical Mechanics (second edition),
John Wiley \& Sons, New York, 1987, Sec.~6.6.

\bibitem{lepri}
S. Lepri, R. Livi, A. Politi, Phys. Rep. 377 (2003) 1.

\end{thebibliography}



\end{document}